# Differential Coded Aperture Single-Snapshot Spectral Imaging

J. Hlubuček[1,2,*], J. Lukeš[1,2], J. Václavík[1,2], K. Žídek[1]

[1]Research Centre for Special Optics and Optoelectronic Systems (TOPTEC), Institute of Plasma Physics of the CAS, Prague, Czech Republic
[2]Faculty of Mechatronics, Informatics, and Interdisciplinary Studies, Technical University of Liberec, Liberec, Czech Republic
*Corresponding author: hlubucek@ipp.cas.cz

**We propose a novel concept of differential coded aperture snapshot spectral imaging (D-CASSI) technique exploiting the benefits of using {-1,+1} random mask, which is demonstrated by a broadband single-snapshot hyperspectral camera using compressed sensing. To double the information, we encode the image by two complementary random masks, which proved to be superior to two independent patterns. We utilize dispersed and non-dispersed encoded images captured in parallel onto a single detector. We explored several different approaches to processing the measured data, which demonstrates significant improvement in retrieving complex hyperspectral scenes. The experiments were completed by simulations in order to quantify the reconstruction fidelity. The concept of differential CASSI could be easily implemented also by multi-snapshot CASSI without any need for optical system modification.**

Hyperspectral imaging (HSI) denotes imaging, where a spectrum is recorded for each pixel of the image. It is a very useful technique for a broad range of samples – for instance, in the infrared (IR) spectral region, light makes it possible to remotely sense the chemical composition owing to specific absorption fingerprints of each chemical compound. Since the acquired dataset is a 3D datacube consisting of many stacked 2D images, HSI inevitably collects a large amount of data. Processing the datacube is very demanding for computation power, acquisition times are usually very lengthy, and the HSI requires a high intensity of light. Moreover, in the IR region, there is a need for special optical materials and IR array detectors. A possible solution to this problem is using a compressed sensing method called Coded Aperture Snapshot Spectral Imaging (CASSI), which makes it possible to compress a 3D hyperspectral scene in a single instant onto a 2D detector and then retrieve the 3D information back thanks to a reconstruction algorithm, such as TwIST [1]. The core of the method lies in encoding a measured scene with a binary random mask pattern which is then spectrally sheared and captured onto a detector. However, since the basic CASSI method relies on a single snapshot, the data compression ratio is immense, making the reconstruction of complex datacubes very problematic. Therefore, an extension of this method is on the spot.

In recent years there have been efforts to decrease the compression ratio by numerous means [2-9]. One of the promising enhancements is (i) acquiring multiple snapshots of the same scene using different random mask patterns [2-4] or (ii) capturing a non-diffracted image of a scene and using this knowledge in the reconstruction [5-9]. However, by (i) the CASSI method requires some advanced modulators due to the need for changing the random mask pattern, and by (ii) a second camera is often needed.

A way to avoid using a second detector could be utilizing a grating in the CASSI system and consequently capturing a zero-order of diffraction, i.e., a non-diffracted image of the scene, next to the first-order – a spectrally sheared image of the scene, on the same detector [10]. Nevertheless, this approach itself does not provide sufficient reconstruction fidelity for real-life HSI in a broad spectral range.

In this letter, we demonstrate a novel approach to obtaining more information about the measured scene in the CASSI technique. The scene is imaged by double-lens and subsequently encoded via two random binary masks. A diffraction grating provides us with both the first-order diffraction image (standard CASSI information) and the zero-order diffraction image, i.e., the spectrally integrated image. We show that the doubled information is a promising way to improve the reconstruction quality without making the optical setup more complex. Moreover, by smart design of the random mask patterns, we are able to improve quality even further.

The imaging of a scene is done with two lenses cut into a rectangular shape (size 10×50 mm, f=100 mm), which were glued together along their long side. The double lens combined with a field lens projects the measured scene into two identical images encoded by two different random masks (64×64 px). The encoded images propagate through the system depicted in Fig. 1A and are captured on the detector above each other. M denotes random mask, L refers to lenses, P and G are prism and grating, respectively, and D are doublets. A detector with resolution 2056x2464 px was used. However, a detector with approximately 145x275 px would be sufficient regarding the mask size and resolution. It is worth noting that our system is also able to capture a non-diffracted image of the scene (zero-order – ZO) on the same detector, which provides us with more information about the measured scene without the need to split an incoming light as in standard extension of the CASSI method [5-9].

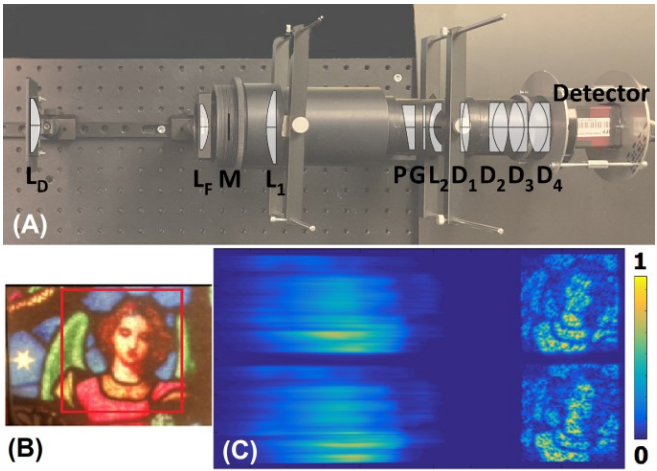

Fig. 1. (A) Scheme of the optical setup. M=mask, L=lens, P=prism, G=grating, D=doublet (see Supplementary Information for more details). (B) Photo of the measured scene. Red square marks the imaged area. (C) Double mask CASSI detector image; first-order of diffraction on the left, zero-order of diffraction on the right. Note that the same color bar applies to all figures.

The two random masks (patterns) encoding the scene might be, in principle, entirely independent and random. However, our simulations prove that it is beneficial to use two complementary masks – see Supplementary Information (SI) for detailed information. Complementary masks approach was used previously for color-coded mask [11]. For a single mask in a standard CASSI system, 50 % of the information is lost at the pixels, where the random mask binary information is 0. Therefore, using two random masks that are complementary to each other, i.e., on the positions of 1's in the first mask are 0's in the second mask and vice versa, we are guaranteed to acquire the information from all the pixels of the scene. It decreases the compression ratio but without the need for the second detector as in [5-9].

In the optical setup, we put stress on a simple construction of the CASSI camera using a minimal number of optical elements. The resulting device is relatively compact and uses a concentric mounting. This approach is kept due to the vision of using an analogous device in the IR regime, where the optical element fabrication and alignment are significantly more challenging than in the VIS region. As a result, our CASSI setup suffers from optical aberrations, which need to be overcome.

Owning to the fact that we have two complementary spectrally sheared images, we can approach the doubled information in several ways. The basic one would be to simply consider the two images separately, as it is done in the multi-frame CASSI extensions [2-3]. We denote this approach as *Double*. Another approach, abbreviated as *Diff*, is to calculate a difference of the detected images, which simulates a measurement with a random mask pattern consisting of ±1s, i.e., a differential image of the two masks. Note that from the CS theory, there is a qualitative difference between the {+1,-1} Bernoulli matrices and the {+1,0} Bernoulli matrices in their compressed sensing performance. It has been proven that CS algorithms work better for mask {+1,-1} [12].

Finally, we also used the approach labeled as *Diffsum*, where we calculated with two images, where one is a difference, and the other one is a sum of the two traces. While we seemingly gain no benefits from using *Diffsum*, the difference and the sum of the two images might be beneficial. It provides us with more information about the image intensity magnitude and improves the reconstructions, as we discuss later.

In Table 1, you can see an overview of the labels. Letter A denotes image corresponding to the upper first-order image, B represents the lower first-order image, D is a detector image, and the final detector image D'= [D ZO], where ZO is a sum of upper and lower zero-order images. Detector image D' is fed to the reconstruction algorithm, whose core is TwIST, which transforms it into a datacube with 123 spectral channels. For the detailed description of data processing, see SI.

**Table 1. Different approaches to data processing**

| Approach | Matrix notation |
|---|---|
| Single | D=[A] |
| Double | D=[A; B] |
| Diff | D=[A-B] |
| Diffsum | D=[A-B; A+B] |

We measured various scenes by using our CASSI-based camera

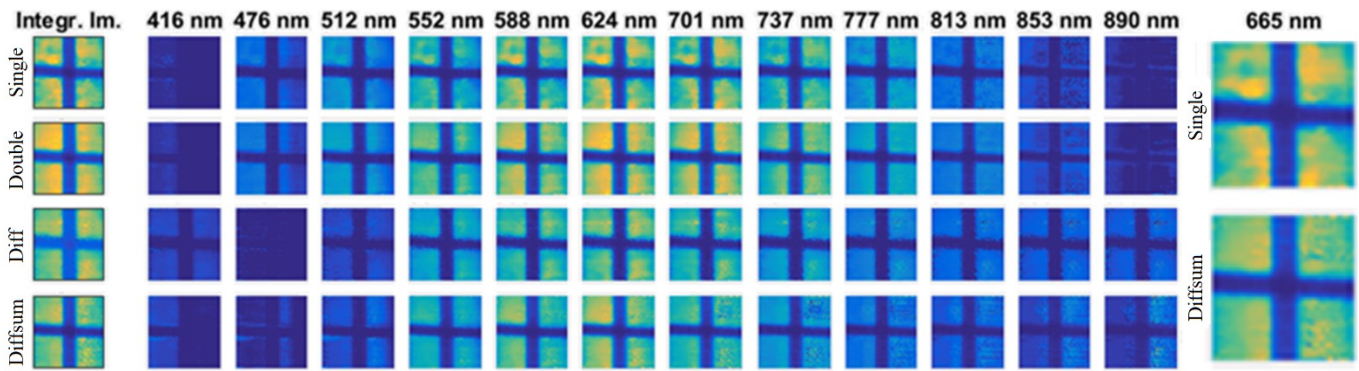

Fig. 2. Reconstructed spectrally integrated image and individual spectral slices of an opaque cross illuminated with broadband light by using four processing approaches (see Table 1). Two spectral slices of *Single* and *Diffsum* were enlarged for comparison.

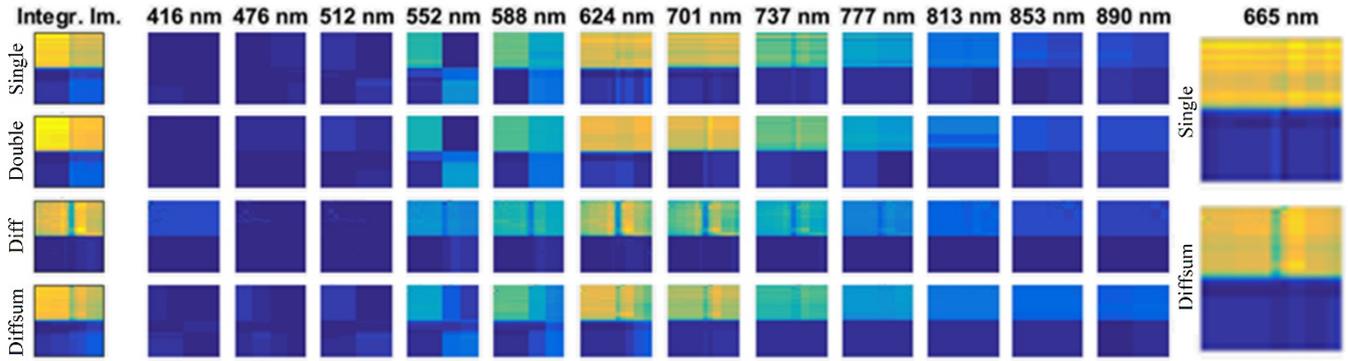

Fig. 3. Reconstructed spectrally integrated image and individual spectral slices of four color filters illuminated with broadband light by using four processing approaches (see Table 1). Two spectral slices of *Single* and *Diffsum* were enlarged for comparison.

and carried out the hyperspectral datacubes reconstructions to test the camera performance.

In Fig. 2, there is a reconstructed scene of an opaque cross evenly illuminated by a white light. We present selected spectral slices for different approaches – compare the lines. There are also two spectral slices of the *Single* and *Diffsum* approach magnified for better comparison. It can be seen that for such a simple scene, the approaches provide us with a similar quality of image reconstruction. However, the intensity distribution in reconstructed spectral slices of *Single* is less homogenous. In addition, all approaches except *Diffsum* leave residual intensity in the region below 500 nm, where the incoming light was cut off by filter OG-515 (see the corresponding spectral slices or Fig. S3 in SI). The zero spectral intensity in this spectral region is, therefore, a useful measure of the spectral reconstruction quality.

While the scene in Fig. 2 consists of a single spectral shape modulated in intensity, Fig. 3 depicts the reconstruction of the scene with four color filters illuminated by a broadband light source. Each quadrant, therefore, featured an entirely different spectrum. *Single*, *Double,* and *Diffsum* were able to accurately recreate the original filters in corresponding quadrants, while *Diff* struggles to reconstruct the green and blue filters in the bottom two quadrants (see spectral slices 552 and 588 nm). This is caused mainly by the aberrations in our system, which are, moreover, slightly different for the upper and lower image. Therefore, the image intensity within one line of a random mask leaks into the neighboring line and distorts the reconstruction. The summed image, which provides a guideline about the actual local intensity on the detector, is then able to compensate for this problem in the *Diffsum* approach. It is worth noting that *Single* or *Double* reconstruction cannot reconstruct well the onset of the yellow filter spectrum (see Fig. S4 in SI). Due to the strong signal from the red filter, we can observe that the spectrum tends to follow the red-filter spectrum in these cases, which is even more prominent with *Diff*. By contrast, the *Diffsum* reconstruction can distinguish between the two aforementioned filters better (see the enlarged spectral slices in Fig. 3). While the benefit of the Diffsum approach for the image quality is not so prominent here, the ability to discern different spectral features is highly improved.

Finally, in Fig. 4, we depict a reconstruction of a complex scene with many varying spectral regions. It is possible to notice the improvement in the spatial quality of the reconstructed images when we extend into the double mask approaches. Nevertheless, even for the *Diffsum* case, we attain only qualitative agreement between the reconstructed and measured spectral shapes of the individual regions.

So far, we have discussed the results qualitatively. Hereafter, we will focus on quantification of the benefits connected to using two complementary masks. It was not possible to quantify the reconstruction quality of the experimental data by residuals from the detector image because we observed that we were able to obtain low residual metrics even for a reconstruction that clearly did not match the original scene. Hence, we created a set of artificial datacubes and detector images faithfully simulating the real detected images by a careful analysis of the aberrations present in our system (see SI for details). We also created a few synthetic detector images using CAVE database data [13] as a template.

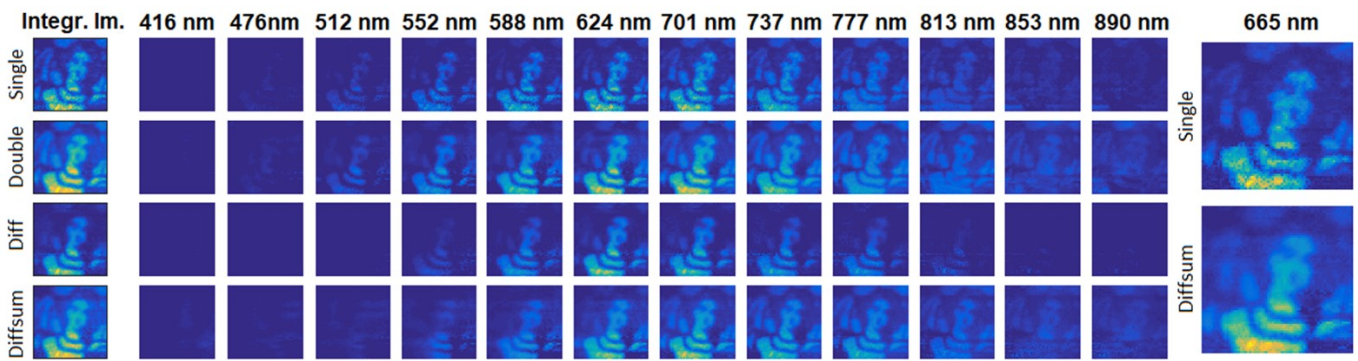

Fig. 4. Reconstructed spectrally integrated image, individual spectral slices, and a spectrum of stained glass foil illuminated with broadband light by using four processing approaches (see Table 1). Two spectral slices of *Single* and *Diffsum* were enlarged for comparison.

Comparison between the original and reconstructed datacube was carried out by the peak signal-to-noise ratio (PSNR), spectral angle mapper (SAM) method for finding the spectral match, and the structural similarity (SSIM) index for measuring the spatial similarity. SAM and SSIM values can be found in SI (Table S3).

Firstly, we evaluated the effect of using two complementary random masks. We compared it to the situation where we use two random masks not related to each other. We consistently attained a higher reconstruction fidelity for the complementary masks (see results in Table S2 in SI). We ascribe it to the fact that we obtain non-dispersed image without any pixels missing for the complementary masks, which in turn improves the reconstruction quality.

Secondly, we applied the simulations on the used complementary masks, where we used the four approaches listed in Table 1. Table 2 provides an overview of the results achieved during the reconstruction for each case characterized by PSNR. We point out that PSNR was calculated by scaling the whole datacube by a single factor, i.e. not by slice-by-slice comparison. In accordance with the real data reconstructions, it proves that the double mask approach *Diffsum* surpass *Single*, while it turned out that *Diff* was providing the worst results for some scenes. We acribe this decrease in the reconstrcution quality to the huge uncertainty in magnitude, which arise when two similar datasets, which are shifted with respect to each other, are subtracted. However, in the case of *Diffsum* the information about the magnitude is still present in sum of the snapshots. This combines the differential character of the random mask, while it retain the information about the image intesity scaling.

For the experimental data, the difference between the modes is more prominent for the complex scenes, while the simple scenes feature a similar image quality. Yet, the fidelity of the reconstructed spectra is improved even for the simple scenes.

**Table 2. Reconstruction results for different scenes**

| Approach | Scene A (cross) | Scene B (filters) | Scene C (feathers) |
|---|---|---|---|
| | PSNR | PSNR | PSNR |
| Single | 17.78 | 21.78 | 21.29 |
| Double | 18.79 | 21.90 | 21.66 |
| Diff | 16.26 | 21.35 | 19.90 |
| Diffsum | 19.67 | 22.11 | 21.85 |

In conclusion, we demonstrated a simple optical setup for a single-snapshot double-image CASSI system on a broad range of 400-900 nm with 123 spectral slices. Our experimental results, confirmed by a set of simulations, show that capturing two images of the same scene encoded by different random masks is superior to the standard approach, i.e., we gain better reconstruction quality. Furthermore, owing to the uniqueness of our system, where we use both zero- and first-order images, we are able to set the initial guess of the datacube very close to the measured datacube, which in turn decreases the number of iterations needed for the reconstruction.

We also performed a comparison of artificial data reconstructions between complementary and non-complementary masks, which confirms that using two complementary masks provides us with more information and, therefore, better reconstruction quality. Hence, our system works like a differential CASSI (D-CASSI) method, where we are able to utilize a random mask consisting of {-1,+1} pixels.

From the selected approaches to the measured data, the best one and, at the same time, the most robust is *Diffsum*, which works well, particularly with the aberrated imaging system. Note that the optical setup was not optimized for the double mask approach, and therefore, the reconstruction quality could be further improved by limiting the aberrations in the system.

In summary, with a simple adjustment of the system without increasing its complexity, one can obtain more information about the measured scene, improve the compression ratio and the reconstructed image quality. Moreover, CASSI systems utilizing spatial light modulators (SLMs), such as DMDs, for generating random mask patterns could benefit from our proposed approach as it can be straightforwardly implemented simply by generating complementary patterns one after another without any need for optical system modification. However, capturing two or possibly multiple images of the same scene encoded by different random masks simultaneously is a way to avoid using SLMs and retain the CASSI main advantage – single-snapshot.

**Funding.** The Ministry of Education, Youth and Sports ("Partnership for Excellence in Superprecise Optics," Reg. No. CZ.02.1.01/0.0/0.0/16_026/0008390); and the Student Grant Scheme at the Technical University of Liberec through project nr. SGS-2019-3054; and project nr. SGS-2021-3022.

**Disclosures.** The authors declare that there are no conflicts of interest related to this article.

**Data availability.** Data underlying the results presented in this paper are not publicly available at this time but may be obtained from the authors upon reasonable request.

See Supplementary Information for supporting content.